\documentclass[prb,aps,twocolumn,showpacs]{revtex4}
\usepackage{graphicx,latexsym}
\usepackage{dcolumn}
\usepackage{amsmath,amssymb,epsf,bm}
\def\eq#1{Eq.~(\ref{#1})}
\def\fig#1{Fig.~\ref{#1}}
\def\vec#1{{\bf#1}}
\begin{document}
\title{Nonuniversality of the intrinsic inverse spin-Hall effect in diffusive systems}
\author{L.~Y. Wang$^{2}$, A.~G. Mal'shukov$^{1}$, and C.~S. Chu$^{2,3}$}
\affiliation{$^1$Institute of Spectroscopy, Russian Academy of
Sciences, 142190, Troitsk, Moscow oblast, Russia \\
$^2$Department of Electrophysics, National Chiao Tung University,
Hsinchu 30010, Taiwan \\
$^3$National Center for Theoretical Sciences, Physics Division,
Hsinchu 30043, Taiwan}

\begin{abstract}
We studied the electric current induced in a two-dimensional
electron gas by the spin current, in the presence of Rashba and
cubic Dresselhaus spin-orbit interactions. We found out that the
factor relating the electric and spin currents is not universal, but rather
depends on the origin of the spin current. Drastic distinction has
been found between two cases: the spin current created by diffusion
of an inhomogeneous spin density, and the pure homogeneous spin
current. We found out that in the former case the ISHE electric current is
finite, while it turns to zero in the latter case, if the spin-orbit coupling is represented by
Rashba interaction.

\end{abstract}

\maketitle
\section{Introduction}
The spin-Hall effect (SHE) and the inverse spin-Hall effect (ISHE) can be
observed in two and three-dimensional electron systems with a strong
enough spin-orbit interaction (SOI) \cite{Dyakonov,Hirsh}. Via this
interaction the electric current induces a flux of spin
polarization flowing in the perpendicular direction and vice versa.
These effects take place in metals and semiconductors, where the
spin-orbit interaction arises from impurity scattering, or band
structure effects. Nowadays they are intensively studying
theoretically (for a review see \onlinecite{Engel}) and
experimentally \cite{Kato, Valenzuela}. These phenomena establish an
important connection between spin and charge degrees of freedom that
can be employed in spintronic applications.

Here we will focus on ISHE. This effect is driven by the spin
current which can be produced by different ways. It can be created
by diffusion of an inhomogeneous spin polarization, or can be
induced directly by a  motive force of various nature
\cite{MalshGeneration,Sipe}. In experimental studies the former
method was used  in Refs. \onlinecite{Valenzuela,Wunderlich_ISHE},
while the latter was employed in Refs. \onlinecite{Werake,
Zhao}. From the theoretical point of view there are two quite
distinct mechanisms of ISHE, depending on the extrinsic or intrinsic
nature of SOI in an  electron system. The extrinsic effect is promoted
by the spin-orbit scattering of electrons from impurities \cite{Hirsh}. The intrinsic effect is associated with the spin-orbit splitting of electron energy bands.
This effect has been studied in Ref.\onlinecite{Raimondi}
together with the extrinsic mechanism. A surprising result of this study
is that the finite inverse SHE takes place even in the case of a pure
intrinsic Rashba \cite{Rashba} SOI, while the direct effect has been
shown to vanish in the considered case of a diffusive system \cite{Mischenko}.  A reasonable explanation is
that the Onsager relation between direct (SHE) and reciprocal (ISHE)
effects should not be satisfied, because the spin-current is not conserving. This argument also
means that for ISHE effect the coefficient in the local linear dependence $I_c=\mathcal{C} I_s$ of the charge current density $I_c$ from the spin current density $I_s$ can depend on the source that originally excites $I_s$. In this sense ISHE is not universal. At the same time, SHE is a universal
effect, because the coefficient relating $I_s$ to $I_c$ does not depend on
how the electric current is produced. It can be created, for example,  by
electron diffusion, as well as by an external electric field. The result will be
the same.  It follows  from the gage invariance of the electromagnetic field.
Formally, one obtains the same spin current, independent on whether it is
induced by the scalar electric potential, or  time dependent vector-potential.

In order to demonstrate the non-universality of ISHE we will consider two
kinds of spin-current sources. In the first case, an inhomogeneous spin
polarization  parallel to the $z$-axis creates the
spin flux due to spin diffusion. In the second case, the spin current is
driven by a spatially uniform "electric" fields, such, that the
fields acting on up and down spins have opposite signs. The latter
situation  corresponds to spin current generation mechanisms
suggested in Refs. \onlinecite{MalshGeneration,Sipe}. Our goal is to
show that the factors $\mathcal{C}$ are different in these two situations.
Since our analysis has shown that in the case of the Rashba spin-orbit
interaction $\mathcal{C}=0$ for the source of the second kind, we will consider
the cubic Dresselhaus interaction, as well, and demonstrate that the Onsager
relation holds in this case.

The outline of the paper is as follows. In Sec.II the linear
response equations relating  the spin and charge currents to the
auxiliary fields will be written for a disordered two-dimensional
degenerate (2DEG) electron gas. From this pair of equations the
auxiliary fields can be excluded and linear relations between the
electric and spin currents can be established. In Sec.III this
theory will be applied to the cases with Rashba  (Subsection A) and
Dresselhaus (Subsection B) spin-orbit couplings. The discussion of
results will be  presented in Sec.IV.

\section{Linear response theory}

The Hamiltonian of the electron system has the form
\begin{equation}\label{H0}
H=H_{0}+V ,
\end{equation}
where $H_{0}$ is the unperturbed Hamiltonian of 2DEG, that includes
the electron's spin-orbit coupling and their scattering on randomly
distributed spin-independent elastic scatterers. The spin-orbit coupling has the general form
\begin{equation}
H_{so}=\vec{h_k}\cdot\bm{\sigma},
\end{equation}
where the effective magnetic field $\mathbf{h_k}$ is a function of the
electron momentum $\vec{k}$ and
$\bm{\sigma}=(\sigma_x,\sigma_y,\sigma_z)$ is the vector of Pauli matrices.
In general, $\mathbf{h_k}$ can be generated by the bulk-inversion
asymmetry in the bulk and structure-inversion asymmetry in a
quantum well (QW). \cite{Winkler} The perturbation
term $V=V_{1}+V_{2}$ represents interactions of electrons with the
auxiliary fields. We will consider  two types of fields. The
first one is a slowly varying in time nonuniform Zeeman field $B$
which is directed perpendicular to 2DEG ($z$- direction). The
corresponding interaction Hamiltonian is
\begin{equation}\label{Vs1}
V_{1} = \sigma_z B\,.
\end{equation}
Another interaction is
\begin{equation}\label{Vs2}
V_{2} = \sigma_z \mathbf{k}\cdot\mathbf{A},
\end{equation}
This Hamiltonian contains the uniform  spin-dependent field
$\sigma_z \mathbf{A}$, where $\mathbf{A}$ slowly varies in time.
Such a field induces the spin current by driving in opposite
directions electrons having opposite spins. It can be
created, for example, by applying a time-dependent strain to a noncentrosymmetric semiconductor. Indeed, as known \cite{Pikus} the strain field $u_{xz}$ gives rise to the spin-orbit interaction $\alpha\sigma_z u_{xz}k_x$. Hence, in this case $A_x=\alpha u_{xz}$. Other mechanisms \cite{MalshGeneration,Sipe} of creating homogeneous spin currents can also be presented in a form of  a spin dependent vector potential that is able to drive spins.

 Within the linear-response theory the spin $I^s$ and charge $I^c$ currents of
noninteracting electrons can be written in terms of retarded $G^r_{\vec{k},\vec{k'}}(
\omega)$ and advanced $G^a_{\vec{k},\vec{k'}}(
\omega)$ single-particle Green's functions. Due to impurity scattering these functions are nondiagonal with respect to the wavevectors $\vec{k}$ and   $\vec{k'}$. The linear-response expressions for the currents, as  functions of the frequency $\Omega$ and wavevector $\vec{q}$, at $\Omega \rightarrow 0$ are given by
\begin{widetext}
\begin{eqnarray}\label{SCcurrent}
\mathbf{I}^{s/c}(\Omega,\vec{q})=-i\sum_{\vec{k},\vec{k'}}\int \frac{d\omega}{2\pi}\langle Tr[\left(G^r_{\vec{k},\vec{k'}}(
\omega)-G^a_{\vec{k},\vec{k'}}(
\omega)\right) \vec{j}_{s/c}
G^a_{\vec{k'}+\vec{q},\vec{k}+\vec{q}}(\omega+\Omega)V(\Omega,\vec{q})
n_F(\omega)  +   \nonumber \\
G^r_{\vec{k},\vec{k'}}(\omega) \vec{j}_{s/c} \left( G^r_{\vec{k'}+\vec{q},\vec{k}+\vec{q}}(\omega+\Omega)
-G^a_{\vec{k'}+\vec{q},\vec{k}+\vec{q}}(\omega+\Omega)\right)V(\Omega,\vec{q})
n_F(\omega+\Omega)]\rangle      +\frac{1}{2}\sum_{\vec{k}}n_F(E_{\vec{k}})Tr[\tau_{s/c},\frac{\partial V(\Omega,\vec{q})}{\partial \vec{k}}]_+
\end{eqnarray}
\end{widetext}
where the spin-current and charge-current operators have the conventional form \cite{Rashba_persistent} $\mathbf{j}_{s/c}=(1/2)[\tau_{s/c},\vec{v}]_+$, with $\vec{v}=\vec{k}/m^{*}+\partial
(\vec{h_k}\cdot\bm{\sigma})/\partial \vec{k}$ and $\tau_s=\sigma_z, \tau_c=e$, $n_F(\omega)$ is the Fermi distribution. In the following the low-temperature case will be assumed, so that $n_F(\omega+\Omega)\simeq n_F(\omega)-\Omega\delta(\omega)$. The angular brackets denote averaging over disorder. This averaging will be performed within the semiclassic approximation, according to the standard procedure \cite{agd}, where we will neglect the weak-localization corrections. We will assume that the spatial variations of the external field are slow within the electron mean free path $l$, so that $lq \ll 1 $. This case corresponds to the diffusion approximation, implying the expansion of \eq{SCcurrent} in  powers of $q$. Also the SOI field will be assumed weak enough, so that $h_{k_F} \ll 1/\tau$, where $\tau$ is the mean electron scattering time.

\section{Inverse Spin Hall effect}
\subsection{Rashba SOI}
\begin{figure}
\includegraphics[width=8cm, height=7cm]{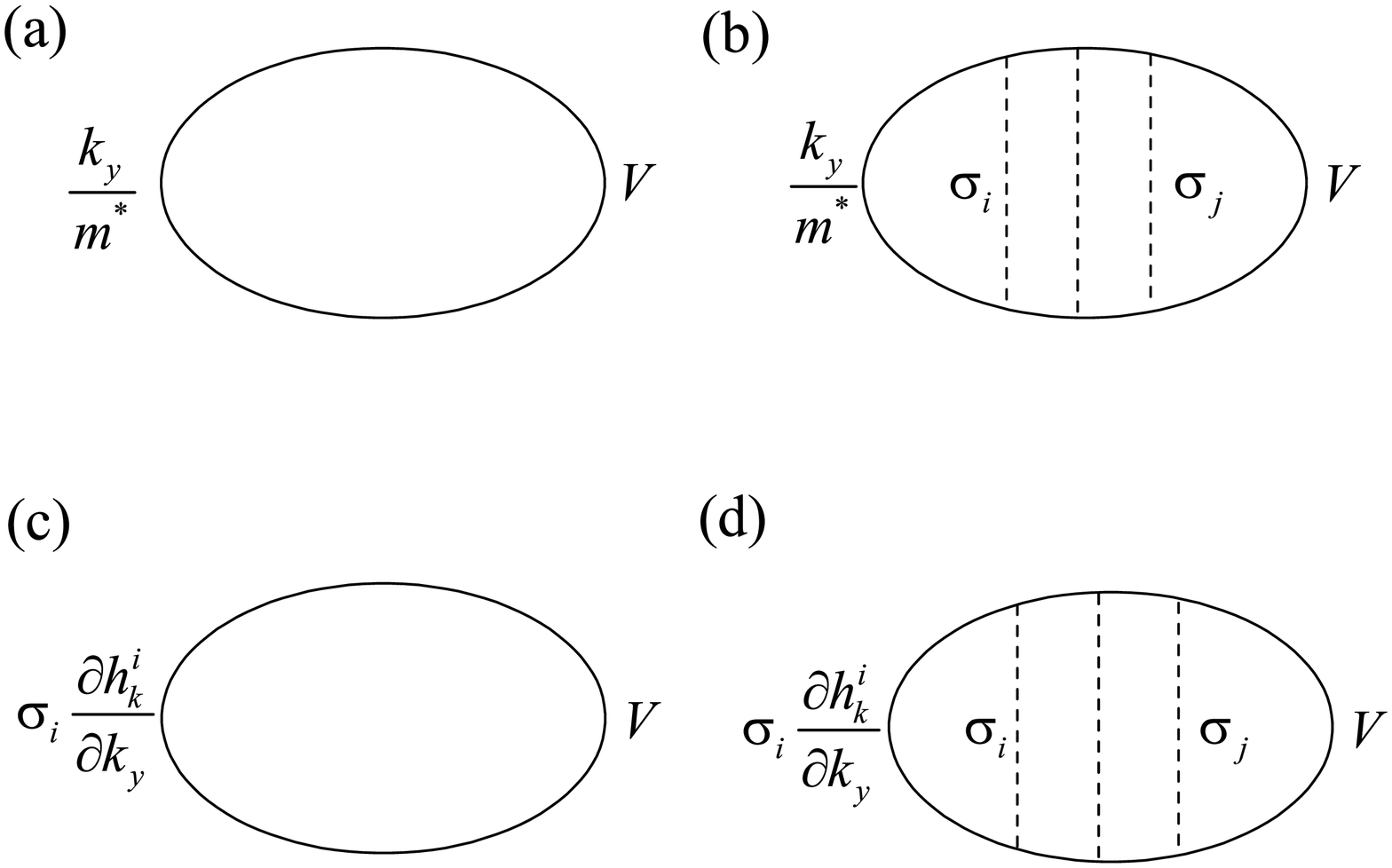}
\caption {The Feynman diagrams for the charge current generated by the
intrinsic spin-Hall effect. The auxiliary field $V$ can be either $V_1$, or $V_2$, where $V_1$ and $V_2$ are defined by \eq{Vs1} and \eq{Vs2}, respectively}\label{fig1}
\end{figure}

Let us first consider ISHE in the case of Rashba spin-orbit interaction, where the spin-orbit field is linear in $k$ and has the form $\vec{h}_\vec{k} \equiv \vec{h}_\vec{k}^R=\alpha\vec{k}\times \hat{z}$. If the auxiliary field is $V_1$, given by \eq{Vs1}, it creates a nonequilibrium and nonuniform in space spin polarization $S_z$.  This  distribution of electron spins relaxes to the uniform state via diffusion, that is accompanied by a pure spin current. When $V(\Omega,\vec{q})$ in \eq{SCcurrent} is represented by $V_1(\Omega,\vec{q})$, the last term in this expression vanishes. Also, the terms containing the products $G^rG^r$ and $G^aG^a$ can be shown to vanish, at least up to linear in $q$ terms. Since in the following the higher-order terms starting from $q^2$ will be ignored, only the products of the form $G^rG^a$   will be retained in \eq{SCcurrent}. We assume that $B$ in \eq{Vs1} varies in $x$-direction, so that $I^c$ is expected to flow in $y$-direction. In Fig. 1 the Feynman diagrams contributing to \eq{SCcurrent}, where $V=V_1$, are shown. The upper (lower) arms in the diagrams denote the impurity averaged functions $G^{r(a)}_{\mathbf{k}}(\omega)=(\omega - E_{\textbf{k}} - \vec{h_k}\cdot\bm{\sigma} \pm i\Gamma)^{-1}$ and the dashed lines depict the random scattering potential correlator  $\langle |U_{\mathbf{k}}|^2\rangle$. For simplicity this correlator will be assumed short-ranged, i.e. independent on $k$, so that $\Gamma=\pi N_F\langle |U_{\mathbf{k}}|^2\rangle \equiv \pi N_F |U|^2=1/2\tau$ is simply a constant. The multiple scattering blocks in diagrams shown at \fig{fig1}(b) and (d) represent processes where the initial electron spin density $S_z$ evolves in the diffusion process to $S_i$.  Since this process is accompanied by the spin precession due to Rashba SOI, $i$ can be either $z$  or $x$, as follows from the spin diffusion equation \cite{Malshdiff} for the spin polarization varying in space along the $x$-coordinate. In general such a diffusion-precession dynamics is represented by the diffusion propagator $D_{ij}(\mathbf{q})$. In the matrix form it can be represented as
\begin{equation}\label{D}
D_{ij}=[(1- |U|^2\Psi/2)^{-1}]_{ij}\,,
\end{equation}
\begin{equation}\label{Psi}
\Psi_{ij}=\sum_{\mathbf{k}}Tr[\sigma_i G^r_{\mathbf{k+q}}(\omega)\sigma_j G^a_{\mathbf{k}}(\omega)]
\end{equation}
Using the above definition, the contribution of all four types of diagrams in \fig{fig1} can be written as
\begin{eqnarray}\label{Icy}
 I^c_y=i\frac{\Omega}{2\pi} B \left(K_{yz}D_{zz}+K_{yx}D_{xz} +\alpha \frac{2\pi N_F}{\Gamma}D_{xz}\right )\,,
\end{eqnarray}
where
\begin{equation}\label{K}
K_{ij}=\sum_{\mathbf{k}}\frac{k_i}{m^{*}}Tr[ G^r_{\mathbf{k+q}}(\omega)\sigma_j G^a_{\mathbf{k}}(\omega)]\,.
\end{equation}
It is easy to see that the diagonal components of $D$ are finite at $q \rightarrow 0$, while the nondiagonal ones vanish as the first power of $q$. Therefore, in the leading approximation the correlator $K$ in the second term of \eq{Icy} must be calculated at $q=0$. Up to the small semiclassic corrections of the order of $(\alpha k_F/E_F)^3$ this correlator is $K_{yx}=-2\pi\alpha N_F/\Gamma$, and the last  two terms cancel each other. At the same time, it is easy to see that $K_{yy}$ is 0 at $q=0$.  Therefore, we did not include the corresponding term $K_{yy}D_{yz}$ into \eq{Icy}. Further, as follows from \eq{K}, the correlator $K_{yz}$ is proportional to  $\vec{h_k}\times\vec{h_{k+q}}$. Therefore, it turns to 0 at $q=0$. In the leading approximation one finds from Eqs.(\ref{Icy}-\ref{K}) that $K_{yz}=-i\pi q \alpha^2 N_F k_F^2/ 2m^*\Gamma^3$  and
\begin{equation}\label{Icyfin}
I^c_y=\frac{\Omega}{2\pi} qBD_{zz} \frac{\alpha^2  k_F^2}{ 4\Gamma^3}
\end{equation}

Our goal is to get an expression of the charge current through the spin current $I^s_x$. Therefore, the next step is to calculate the spin current induced by the perturbation $B\sigma_z$. This current  can be written in the form
\begin{eqnarray}\label{Isx}
I^s_x&=&i\frac{\Omega}{2\pi} B \left( R^x_{zx}D_{xz}+ \nonumber
\right. \\ && \left. D_{zz}\sum _{\mathbf{k}}\frac{k_x}{m^{*}}Tr[
\sigma_z G^r_{\mathbf{k+q}}(\omega)\sigma_z
G^a_{\mathbf{k}}(\omega)]\right )\,,
\end{eqnarray}
where
\begin{equation}\label{R}
R^i_{jk}=\sum_{\mathbf{k}}\frac{k_i}{m^{*}}Tr[ \sigma_j G^r_{\mathbf{k+q}}(\omega)\sigma_k G^a_{\mathbf{k}}(\omega)]\,.
\end{equation}
The second term in the brackets of \eq{Isx} is equal  to $-i\pi N_Fq v_F^2D_{zz}/2\Gamma^2$. This term represents the diffusion spin-current. In its turn the first term is associated with spin precession caused by the Rashba field. It takes a simple form in the case when $q \ll \alpha m^*$, that is when spatial variations of the Zeeman field are slower than spin-density variations caused by spin precession in the SOI field. In this case it follows from \eq{D} that $D_{xz}=|U|^2\Psi_{xz}D_{xx}D_{zz}/2$. A straightforward calculation using Eqs. (\ref{D}),(\ref{Psi}) and (\ref{R}) gives for the first term in the brackets of \eq{Isx} the expression $i\pi N_Fq v_F^2D_{zz}/\Gamma^2$, that is twice larger and has opposite sign with respect to the second term. Finally, from Eqs. (\ref{Icyfin}) and (\ref{Isx}) the charge current becomes
\begin{equation}\label{IcyIsx}
I^c_y=-e\frac{\alpha^2 m^*}{\Gamma}I^s_x
\end{equation}
This result coincides with Ref. \onlinecite{Raimondi}, taking into account that $2\Gamma = 1/\tau$ and that the definition of $I^s_x$ in Ref.\onlinecite{Raimondi} differs by the factor 1/2.

The next example is the charge current  induced in the $y$-direction by  the external perturbation given by Eq. (\ref{Vs2}), where $\mathbf{A}$ is parallel to the $x$-axis. In this case the last term in \eq{SCcurrent} turns to zero, along with the terms containing the products $G^rG^r$ and $G^aG^a$. Further, a simple inspection of diagram (a) in Fig. 1 shows that it is zero at $q=0$. The contribution of other diagrams to $I^c_y$ can be expressed as
\begin{eqnarray}\label{Icy2}
 I^c_y=i\frac{\Omega}{2\pi} A  \sum_i \left(\frac{|U|^2}{2} K_{yi}+\frac{\partial h^i_{\mathbf{k}}}{\partial k_y}\right ) D_{ii}R^x_{iz}\,,
\end{eqnarray}
where the first term corresponds to Fig. 1 (b), while the second
one is given by Fig. 1 (c) and (d). Since $q=0$, only diagonal
components of $D$ enters in \eq{Icy2}. Also, at $q=0$ only $i=x$
must be retained in the sum. As a result, after calculation of
$K_{yx}$,  one can see that that the sum in brackets turns into
zero, up to the small semiclassic corrections of the order of
$(\alpha k_F/E_F)^3$ . Therefore, within the semiclassic
approximation the homogeneous pure spin current can not induce
ISHE. At the same time the spin current created by this source is
finite and is given by the Drude formula
\begin{equation}\label{Is2}
I^s_x=im^{*}\Omega A \frac{v_F^2 N_F}{2\Gamma}\,.
\end{equation}
This expression does not depend on the spin-orbit coupling. The latter enters as a small correction $ \sim h_k^2\tau^2$.

Our calculations in this subsection show that ISHE  is not universal. The induced by this effect electric current is finite, or zero, depending on whether the spin-current  is produced by diffusion of an inhomogeneous spin polarization, or it is a pure uniform spin flux created by an external force of the form Eq. (\ref{Vs2}). The driving force of this sort could be taken into account within the formalism employed in Ref. \onlinecite{RaimondiSU2}. We, however, can not directly see if their expressions for spin and charge currents give, as we expect, vanishing ISHE, because these equations are presented in a rather general form.

\subsection{Dresselhaus SOI}

Although at $V=V_2$ and for SOI given by the Rasha interaction the electric current is zero, we do not expect that the same takes place for a cubic in $k$ Dresselhaus SOI. The reason is that  the spin-Hall effect does not vanish in the latter case \cite{MalshDresselhaus}.  The Dresselhaus SOI field in a quantum well grown along the [001] direction is given by \cite{Eppenga}
\begin{equation}\label{DSOI}
\begin{aligned}
&h_\vec{k}^{x}=\beta k_x(k_y^2-\kappa_z^2),\\
&h_\vec{k}^{y}=\beta k_y(\kappa_z^2-k_x^2),\\
\end{aligned}
\end{equation}
where $\kappa_z^2$ denotes the operator $-\left(\partial/\partial
z\right)^2$ averaged over the lowest subband wave function. Since $h_{\vec{k}}^{i}$ is a nonlinear function of $\mathbf{k}$, $\nabla_{\mathbf{k}} h_{\vec{k}}^{i} $ entering into \eq{Icy2} is not a constant. Therefore \eq{Icy2} has to be modified. Denoting by a bar the average $\overline{\nabla_{\mathbf{k}} h_{\vec{k}}^{i}}$ over the Fermi surface, the modified expression for the current can be written in the form
\begin{eqnarray}\label{Icy2mod}
  I^c_y&=&i\frac{\Omega}{2\pi} A \left[ \sum_i \left(\frac{|U|^2}{2} K_{yi}+\frac{\overline{\partial h^i_{\mathbf{k}}}}{\partial k_y}\right ) D_{ii}R^x_{iz}+ \nonumber \right.\\
  &+& \left. \frac{|U|^2}{2}\sum_{i} \Theta_{i}D_{ii}R^x_{iz}+\Phi \right]\,,
\end{eqnarray}
where
\begin{equation}\label{theta}
 \Theta_{i}=\sum_{j\mathbf{k}}\left(\frac{\partial h^j_{\mathbf{k}}}{\partial k_y}-\frac{\overline{\partial h^j_{\mathbf{k}}}}{\partial k_y}\right)Tr[\sigma_j G^r_{\mathbf{k}}(\omega)\sigma_i G^a_{\mathbf{k}}(\omega)]
\end{equation}
and
\begin{equation}\label{phi}
\Phi=\sum_{ij\mathbf{k}}k_x\left(\frac{\partial h^i_{\mathbf{k}}}{\partial k_y}-\frac{\overline{\partial h^i_{\mathbf{k}}}}{\partial k_y}\right)Tr[\sigma_j G^r_{\mathbf{k}}(\omega)\sigma_z G^a_{\mathbf{k}}(\omega)] \,.
\end{equation}
It is easy to see that the first term in \eq{Icy2mod} turns to zero, similar to  \eq{Icy2} in the Rashba case. However, other two terms are finite, while they vanish for Rashba SOI, as well as for any other SOI which depends linearly on $\mathbf{k}$. Taking SOI in the form of \eq{DSOI}, from definitions (\ref{theta}), (\ref{phi}),  (\ref{R}) and (\ref{D})-(\ref{Psi}) one obtains at $q=0$
\begin{eqnarray}
&&R^x_{yz} =  - 2\pi  \frac{N_F}{\Gamma ^2 }\overline{h_{\mathbf{k}}^x k_x}\,\,\,,\,\,\,R^x_{xz}=0\,\,\,,\,\,\,  \nonumber  \\
&&\Phi =-2\pi  \frac{N_F}{\Gamma ^2 }\left(\overline{\frac{\partial h_{\mathbf{k}}^y }{\partial k_y}h_{\mathbf{k}}^x k_x}-\overline{\frac{\partial h_{\mathbf{k}}^x }{\partial k_y}h_{\mathbf{k}}^y k_x}- \overline{\frac{\partial h_{\mathbf{k}}^y }{\partial k_y }}\overline{h_{\mathbf{k}}^x k_x}\right) \nonumber \\
&&\Psi _{xx}  =\Psi _{yy}= \frac{2\pi N_F}{\Gamma } - \frac{\pi N_F}{\Gamma ^3 } \overline{h_{\mathbf{k}}^2}   \nonumber \\
&&D_{xx}=D_{yy}  =
\frac{2\Gamma^2}{\overline{h_{\mathbf{k}}^2 }}.
\end{eqnarray}
Since only $R^x_{yz}$ is finite in \eq{Icy2mod}, one has to calculate $\Theta_y$. From \eq{theta} it can be expressed as
\begin{equation}\label{theta2}
\Theta_y =\pi  \frac{N_F}{\Gamma ^3 }\left(2\overline{\frac{\partial h_{\mathbf{k}}^x }{\partial k_y}h_{\mathbf{k}}^x h_{\mathbf{k}}^y}-2\overline{\frac{\partial h_{\mathbf{k}}^y }{\partial k_y}h_{\mathbf{k}}^{x2}}+\overline{\frac{\partial h_{\mathbf{k}}^y }{\partial k_y}}\overline{h_{\mathbf{k}}^{2}}\right)
\end{equation}
Collecting all together one obtains from \eq{Icy2mod}
\begin{widetext}
\begin{equation}\label{Icyfin2}
  I^c_y=ieA\Omega \frac{N_F}{\Gamma ^2}\left [2 \frac{ \overline{h_{\mathbf{k}}^x k_x}}{\overline{h_{\mathbf{k}}^{2}}}\left( \overline{\frac{\partial h_{\mathbf{k}}^y }{\partial k_y}h_{\mathbf{k}}^{x2}}-\overline{\frac{\partial h_{\mathbf{k}}^x }{\partial k_y}h_{\mathbf{k}}^x h_{\mathbf{k}}^y}\right) +\overline{\frac{\partial h_{\mathbf{k}}^x }{\partial k_y}h_{\mathbf{k}}^y k_x}- \overline{\frac{\partial h_{\mathbf{k}}^y }{\partial k_y}h_{\mathbf{k}}^x k_x}\right]
\end{equation}
\end{widetext}
This electric current can now be expressed through the spin current. The latter is induced by the time-dependent "vector potential" $A$ in (\ref{Vs2}) and is given by (\ref{Is2}). Denoting by $Q$ the expression in the square brackets of \eq{Icyfin2}, one obtains
\begin{equation}\label{IcyIsx2}
I_y^c=\frac{2eQ}{\Gamma m^* v_F^2}  I_x^s\,.
\end{equation}
Taking into account that $Q\propto h_{\mathbf{k}}^2$ it easy to see that the  charge-to-spin current ratio is of the same order of magnitude as in the case considered in Subsection A, \eq{IcyIsx}, provided that the Rashba and Dresselhaus interactions are comparable in their strengths. One more useful relation can be obtained by using the expression for the spin-Hall conductivity derived in Ref. \onlinecite{MalshDresselhaus,Malshstrip}. This conductivity can be written as $\sigma_{SH}=eN_F Q/\Gamma^2$. Expressing $Q$ in \eq{IcyIsx2} through $\sigma_{SH}$, and writing the electric conductivity in the form of the Einstein relation $ \sigma = 2N_F D$ we find
\begin{equation}\label{IcyIsx3}
I_y^c=\frac{ \sigma_{SH} }{\sigma} I_x^s
\end{equation}
On the other hand, the spin current induced by the spin-Hall effect is given by $I_x^s= \sigma_{SH}E$, where $E$ is the electric field in the $y$-direction. Writing it as $E=I_c^y/\sigma$ we arrive to $I_x^s= \sigma_{SH}I_y^c/\sigma$. This equation, together with \eq{IcyIsx3} establish Onsager relations between spin and charge currents.

\section{Conclusions}

Our analysis shows that the proportionality coefficient in the linear relation between the electric and spin current densities in the inverse spin-Hall effect depends on the origin of the spin current. Therefore,
it is not possible to introduce a universal parameter that determines a charge to spin current  response. This non-universality  is most clearly seen  in the case of Rashba SOI, where a pure spin current produced by diffusion of an inhomogeneous spin polarization gives rise to the finite electric current, while the latter is zero when the spin current is induced by a uniform in space force. In this situation, however, the ISHE produces a finite charge current, if SOI is represented by a nonlinear in $k$ Dresselhaus SOI. It is important that in such a case the spin-Hall effect and ISHE obey the Onsager relation for coefficients relating the spin and charge currents.

It should be noted that calculated above expressions for the spin and charge currents are related to local current densities, while experimentally measured are total electric currents, or electric potentials that are responses not to local spin currents, but rather to currents that are integrated over some
distance (in 2D transport). For example, due to SOI the spin-current density created by spin diffusion oscillates and decay when the distance $x$ from the spin-injection source is increasing. One has to integrate this current over $x$ to obtain the total electric current induced by ISHE.  Since the relation Eq. \ref{IcyIsx} has the local form it will preserve after such an integration.

\begin{acknowledgments}
This work was supported by Taiwan NSC (Contract No.
100-2112-M-009-019), NCTS Taiwan, and a MOE-ATU grant.
\end{acknowledgments}

\end{document}